\documentclass{iopart}

\usepackage{graphicx}

\begin{document}

%%---- Title of the paper 
\letter{Enhanced phase sensitivity and soliton formation
in an integrated BEC interferometer}

%%---- Authors and affiliations
\author{Antonio Negretti\dag\ddag\P\footnote{Current address: 
Department of Physics and Astronomy University of Aarhus and 
Lundbeck Foundation Theoretical Center for Quantum System Research, 
Ny Munkegade, Building 1520, DK-8000 Aarhus C Denmark} 
and Carsten Henkel\dag}

\address{
\dag\ Institut f\"ur Physik, Universit\"at Potsdam, Am Neuen Palais 10, 
D-14476 Potsdam\\
\ddag\ Dipartimento di Fisica, 
Universit\`a degli Studi di Trento, Via 
Sommarive 14, I-38050 Povo (Trento)\\
\P\ ECT*, Villa Tambosi, Strada delle Tabarelle 286, I-38050 
Villazzano (Trento)\\[1ex]
Dated: 15 September 2004
}

%%---- Abstract 

\begin{abstract}
We study the dynamics of Bose-Einstein condensates in
time-dependent microtraps for the purpose of understanding the
influence of the mean field interaction on the performance of
interferometers. We identify conditions where the non-linearity 
due to atom interactions increases the sensitivity of interferometers 
to a phase shift. This feature is
connected with the adiabatic generation of a dark soliton.
We analyze the robustness of this phenomenon with
respect to thermal fluctuations, due to excited near fields in
an electromagnetic surface trap.
\end{abstract}

%% PACS numbers

\bigskip\

%% Introduction

There is currently a large interest in employing Bose-Einstein
condensates (BECs) of dilute atomic gases in the field of matter
wave interferometry \cite{Berman97,Leanhardt04a}.
The advantages of atom interferometers compared to optical ones are
well-known~\cite{Berman97,Godun01}: greater precision due to the large 
atomic mass, 
sensitivity to vibrations, inertial, and gravitational forces,
access to quantum decoherence and to atomic scattering properties, to 
quote a few.
A promising route towards compact applications are integrated
devices, using for example miniaturized optical 
elements built near nanostructured surfaces \cite{Haensel01,Folman02}.

Atom-atom interactions that play a crucial role in BECs, are usually 
considered as a drawback
for matter wave interferometers because they introduce additional
phase shifts and, more fundamentally, quantum fluctuations and diffusion
of the relative phase between the parts of a spatially separated
BEC, see,
e.g.,~\cite{Milburn97b}--\cite{Menotti01}.
\nocite{Imamoglu97a,Javanainen97b,Moelmer97d,%
Anglin01,Petrov01,Bogoliubov01}
In this paper, we show that an operation mode for a BEC
interferometer exists where atom interactions actually enhance the
phase sensitivity. We study in particular the temporal scheme
proposed in Refs.~\cite{Haensel01,Hinds01a} where a
trapped atom sample is split and recombined by slowly deforming 
the trapping potential. 
A similar setting also describes approximately the flow of a 
condensate through an interferometer with
spatially split 
arms~\cite{Stickney02b}--\cite{Andersson02}. 
Interference is 
usually looked for in the lowest vibrational modes of the recombined trap, 
whose relative populations depend on a phase difference imprinted, 
e.g., during the split phase (see, however, \cite{Andersson02} for a 
multi-mode interferometer with non-interacting atoms).
We show here that when a condensate is recombined,
a grey soliton is formed and starts to oscillate with an amplitude
controlled by the interferometer operation. It is quite interesting
that conversely, the interference phase can also be read out 
by measuring the amplitude of either the soliton oscillation or  
the condensate dipole mode. In
both cases, one achieves a better phase sensitivity with detection
schemes that are relatively simple compared to projective measurements
in the vibrational mode basis.
Finally, the enhanced phase sensitivity of the condensate interferometer
is illustrated by numerical simulations
where the output signal is scrambled due to a fluctuating, random potential. 
This models thermally excited magnetic fields that occur in surface-mounted 
microtraps, and our results allow to identify the noise level 
that can at most be tolerated. 

%% ------------- %% ------------- %% ------------- %% -------------
%% Model

\begin{figure}[htbp]
\centerline{%
\includegraphics[height=3.5cm]{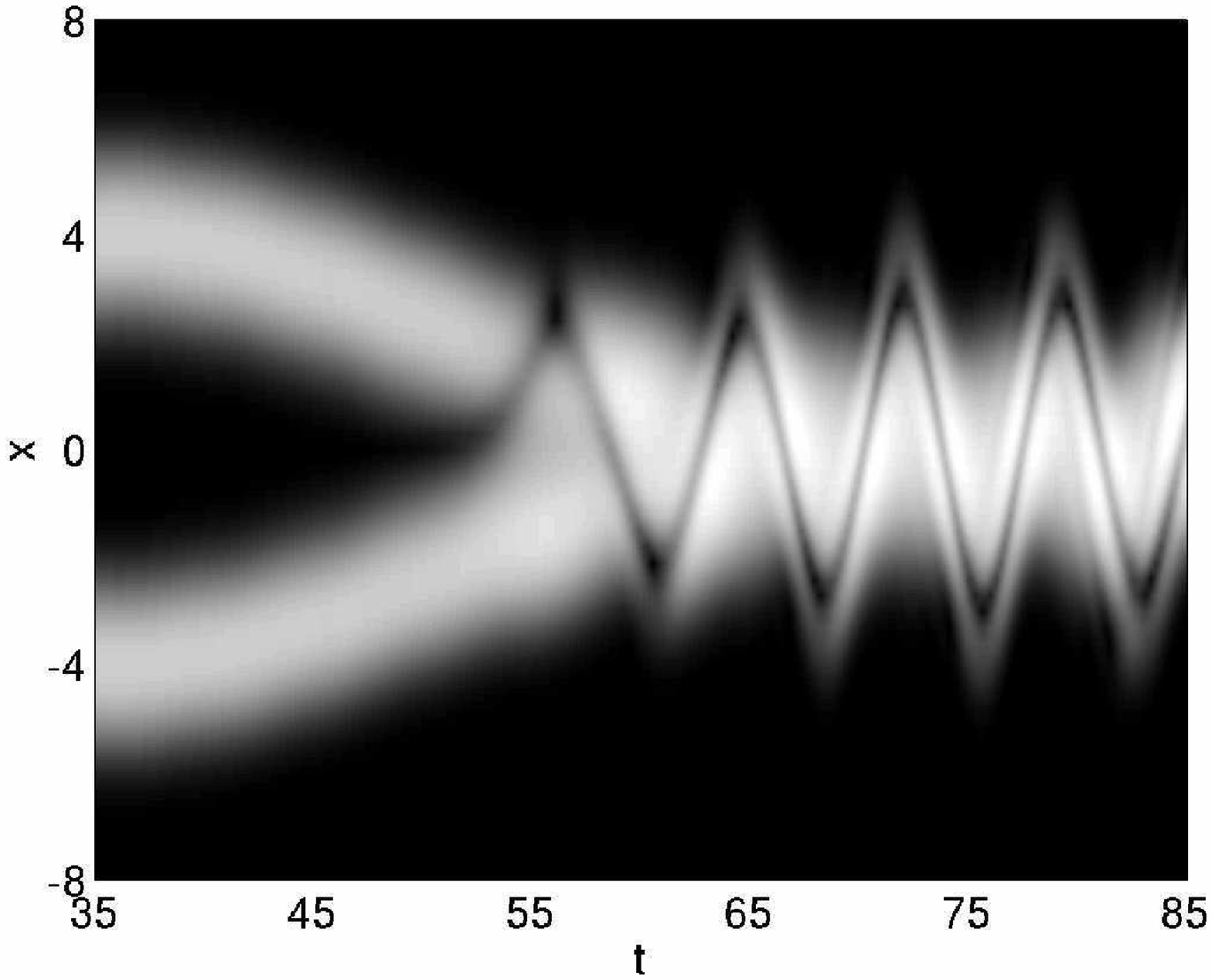}
\includegraphics[height=3.5cm]{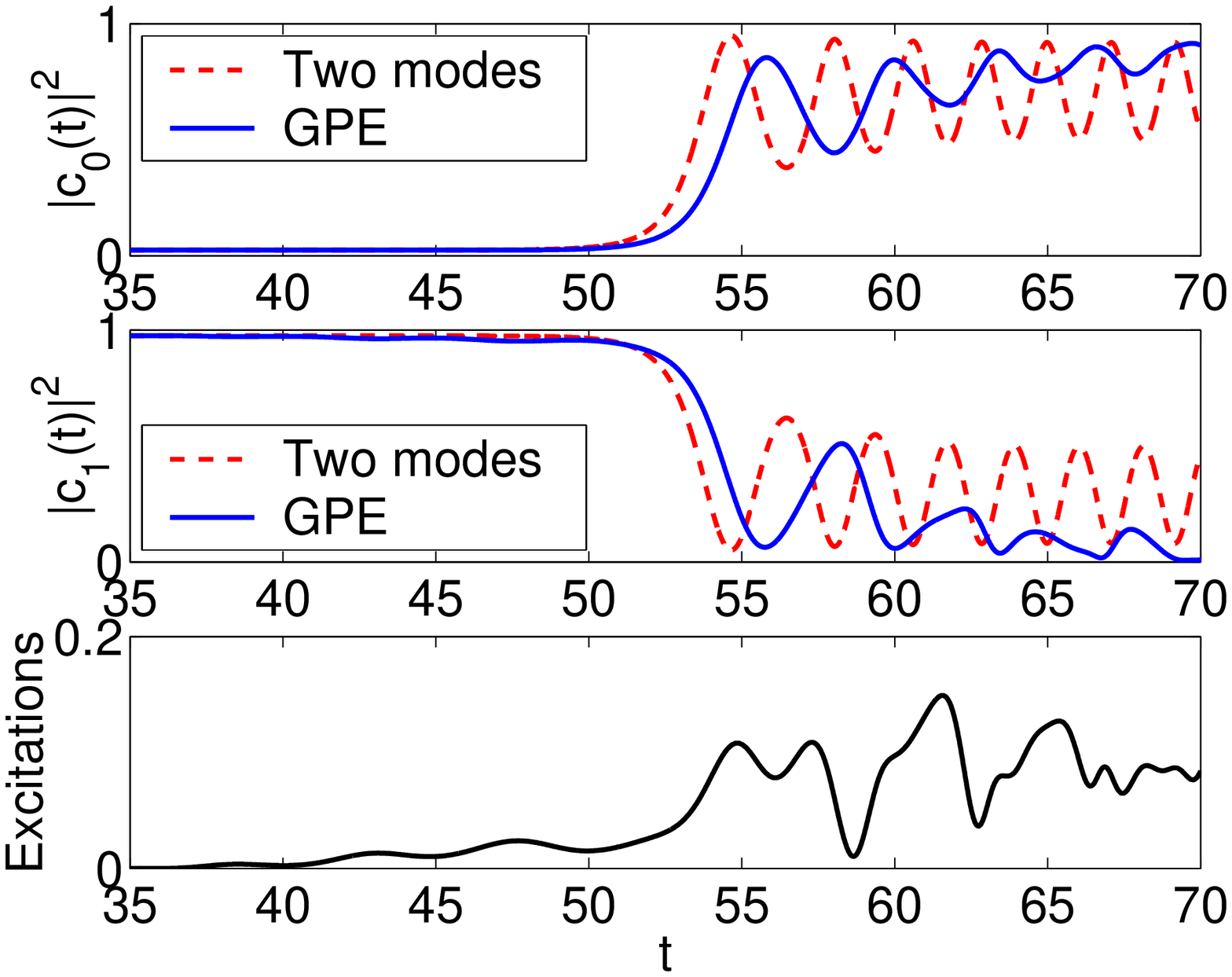}%
}
\caption[]{(left) Recombining the interferometer arms after a phase shift close
to $\pi$ produces a grey soliton oscillating in the harmonic well
(dark zigzag lines), 
as well as dipole and breathing oscillations of the
background condensate (bright). Parameters chosen in the numerical simulation:
$g = 10$, $\Theta = 0.9\,\pi$, operation time $\tau = 70$,
maximum splitting $4a = 8$.
(right) Evolution of the weights $|c_{0,\,1}(t)|^2$ for the first two 
eigenstates of the GPE in the instantaneous trapping 
potential $V( x, t )$ (top and centre) and of the weights of
higher modes (bottom).}
\label{fig:twomodecoef}
\label{fig:recombination}
\end{figure}

In our model for the BEC interferometer, we focus on the 
quasi one-dimensional regime typical for elongated traps,
assume zero temperature
and describe the condensate dynamics along the loosely
bound axis by a one-dimensional order parameter $\Phi( x, t )$
that solves the nonlinear Schr\"odinger or Gross-Pitaevskii
equation (GPE) \cite{Pitaevskii61-Gross63}
\begin{equation}
\label{gpe}
i\hbar\frac{\partial}{\partial 
t}\Phi(x,t)=\left[-\frac{\hbar^2}{2m}\frac{\partial^2}{\partial 
x^2}+V(x,t)+g|\Phi(x,t)|^2\right]\Phi(x,t).
\end{equation}
The potential $V(x, t )$ in Eq.(\ref{gpe}) is harmonic 
for $t \le 0$ and $t \ge \tau$. 
Similar to the setup proposed in Refs.\cite{Haensel01,Hinds01a},
a barrier is adiabatically and symmetrically raised and lowered 
during the operation time $\tau$, splitting and recombining 
the condensate. A relative phase $\Theta$ is imprinted
at time $\tau/2$, when the two wells are sufficiently
separated (negligible tunnelling).  In the following, 
we use harmonic oscillator units defined by 
the initial trap frequency $\Omega$ and adopt 
$u_{l} = \left[\hbar/(m\Omega)\right]^{1/2}$ as unit of length.
We normalize the  order parameter to unity and get a
dimensionless effective interaction constant
\(
g  = 2 N (a_s / u_l )( \Omega_{\perp} / \Omega )
\left( 1 - 1.4603\, a_s / u_{\perp} \right)^{-1}
\)
\cite{Olshanii98},
where $N$ is the number of atoms,
$a_s$ is the three-dimensional scattering length, 
$\Omega_{\perp}$ is the frequency of the transverse (radial) confinement 
and $u_{\perp}$ the corresponding ground state size. We assume a
constant, tight radial confinement and neglect the coupling
between radial and axial excitations.
The numerical integration of Eq.(\ref{gpe}) is done with the 
split-operator algorithm \cite{nrecepiers} whose results we have 
checked for convergence using the conservation of total energy and 
the symmetry under time reversal.
A typical density distribution is shown in Fig.\ref{fig:recombination}(left) 
with a trap potential $V( x, t ) = \frac{1}{2}
\left[x^2-d(t)^2\right]^2 / \left( x^2+d(t)^2 \right)$,
where the well separation is parametrized by 
$d(t)=2a\sin^2(\pi t/\tau)$ for $0 \le t \le \tau$.
The imprinted phase shift is close to $\pi$ so that in the 
ideal gas case, one expects an output state close to the
first excited trap eigenstate. One
clearly sees that a small deviation from this phase shift 
gives a grey soliton that oscillates in the harmonic trap. 

%% ------------- %% ------------- %% ------------- %% -------------
%% Mode analysis

\begin{figure}[htbp]
\centerline{%
\includegraphics[height=3.5cm]{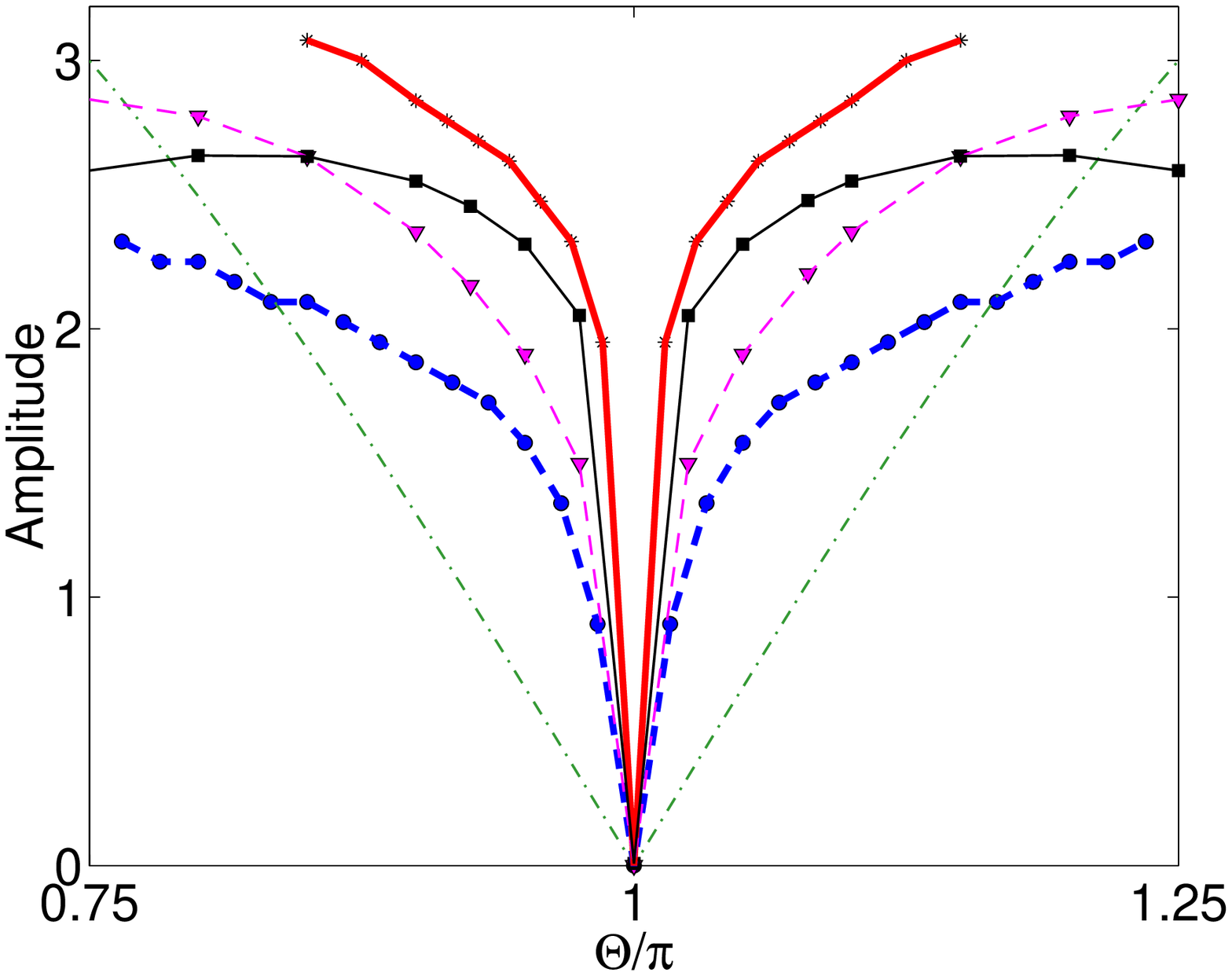}
\includegraphics[height=3.5cm]{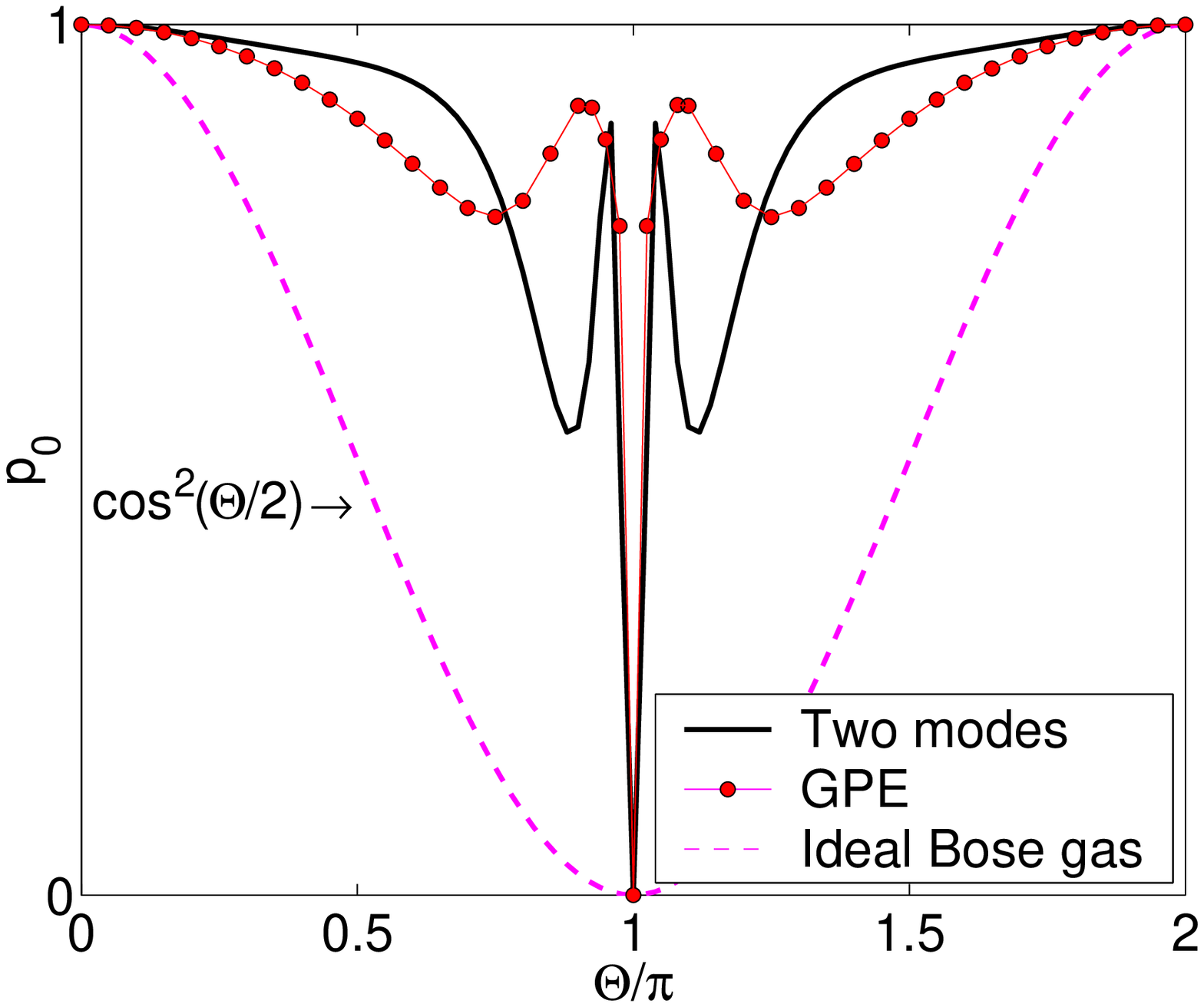}%
}
\caption{(left) Maximum oscillation amplitudes of the mean position 
of the whole condensate and of the soliton position, 
as a function of the phase shift $\Theta$.
For phase shifts very different from $\Theta = \pi$, 
the grey soliton reaches the condensate border in the recombination
stage and escapes; for these values, no results are plotted.
Mean position amplitude ($\times 6$ for clarity): 
$g=0$ (dash-dotted line), $g=5$ (thin dashed line with triangles), 
$g=10$ (thin solid line with squares).
Soliton amplitude: $g=5$ (thick dashed line with dots),
$g=10$ (thick solid line with stars).
(right)
Final population $p_{0}$ of the trap ground 
state, computed by solving the full nonlinear Schr\"{o}dinger equation
(thin solid line with circles, $g=10$), the two-mode model (solid line, $g=10$) of 
Eq.(\ref{eq:c01-mode-evolution}), and the ideal Bose gas (dashed line, $g=0$).}
\label{fig:meanX}
\label{fig:twomodepopu}
\end{figure}

A more quantitative characterisation   
is usually based on the 
final populations $p_{0,1}$ of the lowest trap eigenstates
($k = 0, \, 1$)
\begin{equation}
\label{popu}
p_k = \left|\langle \phi_k | \Phi(\tau) \rangle \right|^2
\equiv
\Big| 
\int\!{\rm d}x \, \phi_k^*( x ) \Phi( x, \tau )
\Big|^2
,
\end{equation} 
where 
$| \Phi(\tau) \rangle$ is the order parameter at time $t=\tau$.
We choose here the convention that $| \phi_{0,1} \rangle$ 
are the ground state and the first 
excited state in the harmonic trap, \emph{including the
nonlinearity}. We illustrate below that this 
improves the accuracy
of a two-mode approximation (see Fig.\ref{fig:twomodepopu}).
Two alternative quantities, that appear simpler to be measured in practice,
are the condensate mean position after recombination
and the centre of the density dip characteristic for the grey soliton 
\begin{eqnarray}
\label{meanpos-solpos}
\langle x \rangle_t = \int\!{\rm d}x \,|\Phi(x,t)|^2 x,
\qquad 
q_t ={\rm min_{local}}|\Phi(x,t)|^2
\qquad 
\mbox{for } t > \tau.
\end{eqnarray}
These quantities show oscillations whose amplitude depends on
the imprinted phase shift $\Theta$, 
as shown in Fig.~\ref{fig:meanX}(left). 
The key observation is that around a complete swapping to the odd 
output state, the curves are much narrower 
compared to the non-interacting case where the dipole mode oscillation
amplitude is $|\sin\Theta|/\sqrt{2}$.
In this sense 
the nonlinearity increases the phase sensitivity of the 
interferometer.%
\footnote{We have checked that the soliton oscillation 
agrees with the effective equation of motion of Ref.\cite{Busch00}
in the Thomas-Fermi limit, i.e.\ for interactions $g \gg 1$.
The difference 
with respect to Ref.\cite{Konotop03} may be only apparent and related 
to the different scaling convention for the dimensionless parameters.}

The populations $p_{0,1}$ show a similar behaviour which can be
understood using a two-mode model similar to those studied in
Refs.\cite{Menotti01,Stickney02b,Stickney03a}. We consider an
expansion of the order parameter
\begin{equation}
    \Phi(x,t) = \sum_{k}c_k(t)\phi_k(x;d(t)),
\end{equation}
where $\phi_k(x;d(t))$ are (real) eigenstates of the GPE
in the double-well potential with $d(t)$ and $g$ fixed.
In the adiabatic approximation and restricting the expansion to the 
lowest even and odd modes $k = 0,\, 1$, the equations for the 
coefficients are 
\begin{equation}
    \label{eq:c01-mode-evolution}
i\,\dot{c}_k = 
\mu_k c_k + g \left( 
2 \mathcal{O}_{01}
- \mathcal{O}_{kk} 
\right) 
|c_{1-k}|^2 c_{k} +
g \mathcal{O}_{01}
c_{1-k}^2 c^*_{k}
\end{equation}
where $\mu_k$ are the chemical potentials of the eigenstates. 
Due to parity conservation, the only nonzero interaction matrix 
elements are
\begin{equation}
\mathcal{O}_{kl} = 
\int\!{\rm d}x\left[ \phi_{k}(x;d)\phi_{l}(x;d) \right]^2
.
\end{equation}
Starting from the ground state $\phi_{0}( x; d(0) )$, adiabatic 
evolution and phase imprint lead to the amplitudes
$c_{0}( \tau/2 + 0 ) = 
\cos(\Theta/2) c_{0}( \tau/2 - 0 )$ and $c_{1}( \tau/2 + 0 ) = 
{\rm i} \sin(\Theta/2) c_{0}( \tau/2 - 0 )$ up to small corrections
due to tunnelling between the wells.
With an exact $\pi$ phase shift, 
$c_{0}(t ) \equiv 0$ for $t > \tau / 2$ as in the non-interacting case.
But for $\Theta \ne \pi$, both modes are nonlinearly coupled.
Solving Eqs.(\ref{eq:c01-mode-evolution}) numerically, we obtain the 
results of Fig.\ref{fig:twomodepopu}(right) for the final population 
$p_0$. 
Comparing to the full solution of the GPE, 
both curves show almost the same behaviour
in the close vicinity of $\Theta = \pi$, and agree qualitatively
for other phase shifts: they both present a maximum and a 
minimum. This differs from the results of \cite{Stickney03a}, 
probably because there the eigenstates of the linear Schr\"{o}dinger 
equation were used for the expansion. 
Moreover, a more accurate agreement could not be expected because
the two-mode approximation is valid only when the low
energy states in the separated wells are weakly modified by the
many-body interactions. 
Estimates available in the literature,
e.g.\ Ref.\cite{Milburn97b}, show that 
the two-mode approximation is limited to an interaction strength
$g \ll 2^{3/4} \sqrt{\pi} \approx 2.98$, given our trap potential.

In Fig.\ref{fig:twomodecoef}(right), we compare the evolution of 
the weights $|c_{0,1}( t )|^2$ between the full numerical solution and 
the two-mode approximation for a phase shift close to $\pi$. Again, 
the approximation reproduces the general features, with quantitative 
differences in the amplitudes and the frequencies of the 
oscillations. 
The curves also illustrate that for superpositions
of instantaneous eigenstates, the adiabatic approximation fails due to 
the nonlinearity: 
the state after the phase imprint---it is almost identical to
the lowest antisymmetric state---turns during the 
evolution into a state with a strong ground state admixture. 
This is due to the presence of an instability in the 
system. More precisely, the excitation spectrum
of the antisymmetric state contains a soft 
mode~\cite{Stickney03a,Fedichev99}.
Solving the Bogoliubov-de Gennes equations, we have found that the
first eigenmode acquires a purely imaginary frequency if the 
splitting $d$ between the wells exceeds some critical value that 
becomes smaller with increasing interactions.\footnote{We were unable
to reproduce quantitatively the imaginary eigenfrequency plotted in 
Ref.\cite{Stickney03a}. This may be related to the different
ansatz for the perturbed order parameter used there.}
The imaginary excitation frequency
leads to an exponential growth 
of the ground state, and this instability is rooted in the
Josephson effect \cite{Reinhard97,Pitaevskii-Stringari01}:  
putting a relative phase different from $\pi$ on the two arms of the interferometer,
one introduces a tunnel current across the barrier that drives
a population imbalance between the two wells. The interactions
in each well subsequently enhance the phase difference, leading
to a runaway effect for the antisymmetric state. The instability 
is weaker at large
splitting because tunnelling becomes exponentially suppressed.
This is consistent with the behaviour of the growth rate found from
the Bogoliubov analysis (see also~\cite{Stickney03a}).
For the chosen
parameters the timescale of the process is actually dominated
by the merging of the two wells that increases Josephson critical 
current. As illustrated in Fig.\ref{fig:twomodecoef}(right),
the main population transfer occurs
around the moment when the barrier between the two wells 
drops below the chemical potential.

%% ------------- %% ------------- %% ------------- %% -------------
%% Dynamics in noisy traps

\begin{figure}[htbp]
\centerline{
\includegraphics[width=8cm,height=4.5cm]{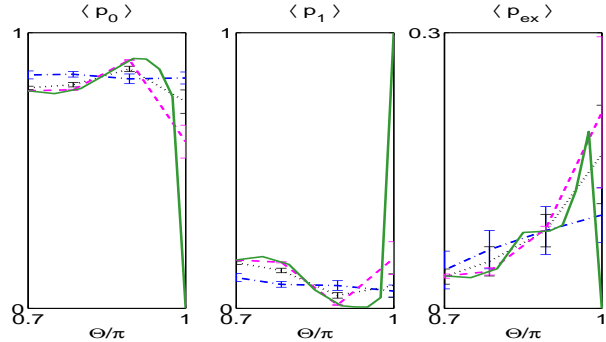}%
}
\caption{%
Average populations $\langle p_{0} \rangle$, 
$\langle p_{1} \rangle$, and $\langle p_{{\rm ex}}\rangle$ 
(higher excitations) in a noisy interferometer for different 
values of the noise power spectral density: 
$\gamma=0.01$ (dash-dotted line), 
$\gamma=0.001$ (dotted line), $\gamma=0.0001$ (dashed line), and 
without noise (solid line). The error bars indicate the statistical
uncertainty of the simulation.}
\label{fig:noiseA}
\end{figure}

The previous results suggest that in a realistic setting, the condensate
interferometer is very sensitive to fluctuations of the trapping
potentials. It has also become clear recently that electromagnetic field
fluctuations are
particularly relevant for integrated atom optics in the neighbourhood 
of material microstructures held at a finite temperature 
(see \cite{Folman02,Scheel04a,HenkelHabil} and references therein).
We have simulated these fluctuations in terms of a random potential that
fluctuates in time and space with spectral characteristics similar
to those of thermal magnetic near fields: white noise and a lorentzian
spatial correlation function~\cite{Henkel03}. 
We focus on the case that the correlation length is smaller than the 
distance between wells. This is realistic for microtraps whose splitting
is larger than the distance to the underlying surface. 
Fig.\ref{fig:noiseA} shows the 
impact of increasing the noise strength for the interferometer output, 
averaged over a statistical sample of realizations.
The steep structures around a $\pi$ phase shift are smoothed,
as expected. It is surprising that this happens already for very weak 
noise: if the scattering rate off the noise potential is denoted 
$\gamma$, significant changes are seen for $\gamma \tau \sim 0.07$ 
already. The instability in the double-well potential thus makes the 
condensate more sensitive to thermal near fields. This differs from
a non-split condensate in a single well, which is more robust in the
presence of noise compared to an ideal gas, as shown in~\cite{Henkel04}.

Throughout this paper, we have used the mean field description
based on the Gross-Pitaevsk\^i equation and neglected phase
fluctuations. At fixed $g$, this can be justified for 
a sufficiently large number of atoms.  
The phase fluctuations in the initial condensate are still small
even at finite temperatures 
$0 < T < T_{\varphi} \approx 
N / ( \Omega_\perp/\Omega + (3 g / (4\sqrt{2}))^{2/3})$ 
(in our units) 
\cite{Petrov01}, using the regime of tight radial confinement.  
Similarly, the phase coherence of the split
condensate \cite{Javanainen97b,Menotti01}
is maintained for operation times
$\tau < \tau_{\rm diff} \approx (2 \sqrt{3} / g)^{2/3}N^{1/2}$, 
following the procedure explained in Ref.\cite{Moelmer97d}.

\smallskip\

A. Negretti acknowledges the financial support of the 
European Union's Human Potential Programme under contract 
HPRN-CT-2002-00304 (FASTNet). A. Negretti thanks C. Menotti, L. P. 
Pitaevskii, and M. Wilkens for stimulating discussions, 
the Institut f\"ur Physik in Potsdam for the friendly hospitality,
and the ECT* for giving the opportunity to use its facilities. 
We thank Th. Busch and N. P. Proukakis for helpful suggestions 
to the manuscript.
This work has been supported by the European Union's IST Programme
under contract IST-2001-38863 (ACQP).

\smallskip\

\end{document}